\documentstyle[aps]{revtex}
\begin{document}

\title{
Appearance of new lines and change in line shape in the IR
spectrum of a NaV$_2$O$_5$ single crystal at the spin-Peierls transition
}
\author{M.\,N.\,Popova\footnote{e-mail: popova@isan.troitsk.ru}
and A.\,B.\,Sushkov}
\address{Institute of Spectroscopy of Russian Academy of Sciences,
142092 Troitsk, Moscow reg., Russia}
\author{A.\,N.\,Vasil'ev}
\address{Physics Department, Moscow State University, 119899 Moscow, Russia}
\author{M.\,Isobe and Yu.\,Ueda}
\address{Institute for Solid State Physics, The University of Tokyo \\
7--22--1 Roppongi, Minato-ku, Tokyo 106, Japan}
\date{\today}
\maketitle

\begin{abstract}
We report the first observation of
new lines in the infrared spectrum of a crystal due to unit-cell
doubling at the spin-Peierls transition (in NaV$_2$O$_5$). The change in the
shape of the spectral lines at the spin-Peierls transition is
recorded for the first time.
A contour characteristic of a Fano resonance is observed above the transition
temperature $T_{\rm sp}$ and the standard symmetric contour is observed
below $T_{\rm sp}$. We attribute this effect to the opening of a gap in the
magnetic-excitation spectrum at the spin-Peierls transition.
\end{abstract}

\bigskip

A number of interesting effects, both of a purely quantum nature and due
to the interaction of the magnetic subsystem with lattice deformations, are
observed in low-dimensional magnetic systems. Specifically, the
magnetic-excitation spectrum of a uniform antiferromagnetic (AF) chain of
Heisenberg spins contains a gap in the case of integer spins and no gap in
the case of half-integer spins.\cite{c1} However, in a real crystal the
interaction of one-dimensional magnetic chains of half-integer spins with a
three-dimensional phonon field results in dimerization of the atoms in the
chains. As a result of such a structural phase transition initiated at the
temperature $T_{\rm sp}$ by a spin-phonon interaction, the magnetic
excitation spectrum acquires a gap that separates the nonmagnetic singlet
ground state from the first excited triplet state.\cite{c2}
A transition of this type is the magnetic analog of the Peierls transition in
quasi-one-dimensional conductors and, for this reason, it is called the
spin-Peierls (SP) transition. The SP transition was first observed in several
organic compounds (see, for example, Ref.~\cite{c2}) and then in inorganic
transition-metal oxides CuGeO$_3$ and NaV$_2$O$_5$ (Refs.~\cite{c4} and
\cite{c5}).

In contrast to their organic analogs, inorganic SP compounds can be obtained
in the form of large, high-quality single crystals. This has made it possible
to expand the arsenal of methods for investigating the SP transition and to
obtain new and fundamental results. For example, for CuGeO$_3$ it is possible
to observe the energy gap ($\Delta$=2\,meV)\cite{c51} directly by the
inelastic neutron scattering method. Magnon-like excitations with an
appreciable dispersion exist below the SP transition temperature
($T_{\rm sp}$=14\,K), while above $T_{\rm sp}$, all the way up to
temperatures close to the temperature of the maximum in the magnetic
susceptibility (60\,K), magnetic fluctuation modes exist in a wide range of
energies.\cite{c52}
Dimerization of the copper ions along chains has been observed at
$T=T_{\rm sp}$, and critical structural fluctuations have been investigated
near $T_{\rm sp}$ (see, for example, Ref.~\cite{c52}).

Since dimerization in magnetic chains at the SP transition is accompanied by
unit-cell doubling in the direction of the chains, the Brillouin zone (BZ)
``folds'' along this direction, and the zone boundary phonons are
transferred to the center of the BZ, in principle making it possible to
observe them at $T<T_{\rm sp}$ in the first-order optical spectra. A
corresponding change in the spin-excitation spectrum at the SP transition
can appear in the ``two-magnon'' spectra. Spectral manifestations of the
spin-phonon interaction in the form of asymmetric lines can also be
expected in SP systems (Fano resonance~\cite{c6}). The effects listed above
have all been observed~\cite{PHM} in the Raman spectra of CuGeO$_3$. In the
present work the appearance of new lines accompanying the SP transition and
Fano resonances were observed in the infrared (IR) spectra of a SP
compound. A change in spectral line shape at the SP transition was
detected.

We investigate the SP vanadate NaV$_2$O$_5$, which belongs to the
orthorhombic system (space group $P2_1mn$) whose structure contains magnetic
and nonmagnetic chains of V$^{4+}$O$_5$ (S=1/2) and V$^{5+}$O$_5$ (S=0)
pyramids oriented along the $b$ axis and alternating with one another in the
$ab$ layers. The Na atoms lie between the $ab$ layers~\cite{c7}.
The magnetic susceptibility at temperatures above 35\,K is described well by
the expression for an antiferromagnetic chain of Heisenberg spins $S$=1/2,
coupled by the exchange interaction ${\cal H} = \sum_i \rm
JS_{\mit i}S_{\mit i+1}, J=560$\,K, and below 35\,K the magnetic
susceptibility decreases isotropically~\cite{c4,c8}. At the same time, new
reflections appear in the x-ray scattering.\cite{c5} The detailed
temperature dependence of the intensity of one reflection gave for the
transition temperature $T_{\rm sp}=35.27\pm0.03$\,K.\cite{c5}
At $T=10$\,K it was possible to determine the propagation vector for a new
superstructure: ${\bf q}=(1/2, 1/2, 1/4)$.

Single crystals, extended along the $b$ axis, with the approximate dimensions
1$\times$4$\times$0.5\,mm along the $a$, $b$, and $c$ axes, respectively,
were obtained in the manner described in~\cite{c4,c5}. The orientation of the
crystals and the lattice constants ($a=11.318, b=3.611, c=4.797$\,\AA) were
determined by the x-ray method. We employed samples on which EPR and
susceptibility measurements had been performed previously~\cite{c8}.
For the IR measurements we have prepared the 13--20\,$\mu$m thick
platelets in the direction of the $c$ axis.
A BOMEM DA3.002 Fourier spectrometer  was used to
measure the transmission spectra in the region 50--400\,cm$^{-1}$ with a
resolution of 0.05--1\,cm$^{-1}$. The samples were placed in a helium
atmosphere at a fixed temperature 5--300\,K in an optical cryostat with 20
$\mu$m thick ``warm'' and ``cold'' mylar windows. To increase the accuracy
of the transmission measurements in the temperature range from 300 to 5\,K,
a special inset in the optical cryostat was fabricated. This insert
partially compensated the thermal contraction of the cryostat column and,
correspondingly, the motion of the sample, and it also made it possible to
register the reference spectrum at any temperature without removing the
sample.

The vibrational spectrum of a NaV$_2$O$_5$ crystal at $T>T_{\rm sp}$ contains
45 fundamental vibrations: $15A_1+8A_2+7B_1+15B_2$. The $A_2$ vibrations are
inactive in IR absorption. In our experimental geometry ({\bf k}$||${\bf c})
$A_1$ vibrations can appear in {\bf E}$||${\bf a} and $B_1$ vibrations can
appear for {\bf E}$||${\bf b}. At room temperature two $B_1$ vibrations
($\nu_{\rm TO}=175, 367$\,cm$^{-1}$) and two $A_1$ vibrations ($\nu_{\rm TO}
= 140$ and 251\,cm$^{-1}$) are seen in the experimental frequency range.
The peak at 140\,cm$^{-1}$ is strongly asymmetric. As the temperature
decreases to 200\,K additional lines become visible at the frequencies 168,
215, and 226\,cm$^{-1}$ ($B_1$) and an asymmetric peak appears near
90\,cm$^{-1}$ ($A_1$).

Fig.\,1 shows the transmission spectra of a 13-$\mu$m thick NaV$_2$O$_5$
sample in two polarizations at temperatures above and below $T_{\rm sp}$.
An intense peak of a $B_1$ vibration and lines with frequencies 168,
215.1, and 225.7\,cm$^{-1}$, which are probably due to the weakly active
$B_1$ vibrations (the nature of these lines needs to be determined more
accurately), are seen in {\bf E}$||${\bf b} polarization at $T=42$\,K against
the interference background in the sample. At $T<T_{\rm sp}$ new lines appear
and grow at the frequencies 101.5, 111.7, 127, 199.7, 234, and 325\,cm$^{-1}$.
The first two lines are vary narrow (about 0.2\,cm$^{-1}$), the third line is
a doublet (126.7 and 127.5\,cm$^{-1}$) consisting of lines which are just
narrow,\footnote{One component of the doublet is saturated, which can distort
the recorded line shape.} and the remaining lines have a width of about
2\,cm$^{-1}$.

A wide absorption continuum is present in {\bf E}$||${\bf a} polarization.
The intensity of its long-wavelength part decreases appreciably at
$T<T_{\rm sp}$. Against the background of this ``bleaching'' of the crystal,
the shape of the peaks at 90 and 140\,cm$^{-1}$ changes from strongly
asymmetric to symmetric. Fig.\,2 illustrates this for the example of the
90\,cm$^{-1}$ peak (the saturation of the
140\,cm$^{-1}$ peak at low temperatures distorts its shape). At
$T>T_{\rm sp}$ this line possesses a shape characteristic for a Fano
type resonance (the resonance between a discrete level and a continuum of
states\cite{c6}). Its width decreases with decreasing temperature while the
shape remains the same.  At $T<T_{\rm sp}$ a narrow absorption line is
observed at frequency 91.2\,cm$^{-1}$.  At $T<T_{\rm sp}$ new lines appear
in the spectrum at the frequencies 101.4, 101.7, 126.8, 127.5, 145.7, 148,
157.2, and 199.5\,cm$^{-1}$.  The lines with frequencies below
199\,cm$^{-1}$ are very narrow (0.2--0.4\,cm$^{-1}$) and the line at
199.5\,cm$^{-1}$ is much wider.\footnote{This line is close to saturation,
and for this reason its width cannot be determined correctly.}

We shall now discuss our experimental data. According to the x-ray scattering
results~\cite{c5} for NaV$_2$O$_5$, at $T_{\rm sp}$ the unit cell doubles in
the directions of the $a$ and $b$ axes and quadruples in the direction of the
$c$ axis. Hence it follows that at $T<T_{\rm sp}$ phonons from the
boundaries of
the BZ along all the directions become IR active. The narrow lines which we
observed to appear at $T_{\rm sp}$ are evidently due to them. Fig.\,3
displays the integrated intensity $I$ of one such line as a function of the
temperature. Since I$\sim \delta^2$, where $\delta$ is the displacement of
the atoms in the lattice at a structural phase transition (the order
parameter), the function $I(T)$ reflects the temperature variation of the
squared order parameter. The accuracy of our measurements is too low for
analyzing the applicability of different theories for describing the
transition under study.

The absorption continuum observed in {\bf E}$||${\bf a} polarization is
apparently of a magnetic nature. On the basis of the data~\cite{c4} on the
temperature dependence of the magnetic susceptibility of NaV$_2$O$_5$, which
has a wide maximum near room temperature, it can be concluded that in a
quasi-one-dimensional AF spin system of NaV$_2$O$_5$ short-range order
persists right up to room temperature, and therefor magnetic
excitations exist.

Wide bands had been observed previously in the absorption spectra of
antiferromagnets. They were attributed to phonon absorption processes
accompanied by the creation of two magnons with equal and oppositely directed
wave vectors {\bf q} (see, for example, Ref.~\cite{c9}). To find the
probability of such a process it is first necessary to examine the
interaction of light with a pair of magnetic ions. The analysis for a pair of
the nearest magnetic ions V$^{4+}$ in NaV$_2$O$_5$ (the symmetry group of the
pair is $C_s$)shows that the first term in the Hamiltonian describing the
interaction of the magnetic system with light makes the largest
contribution~\cite{c9}:
$${\cal H}= \sum_{\mit i} ({\bf E}\cdot{\bf \pi})
({\bf S_{\mit i}}\cdot{\bf S_{\mit i+1}}).$$
Here {\bf E} is the electric field vector of the light wave; the vector
$\bf \pi$ lies in a plane perpendicular to the direction of the spin
chain and its magnitude is proportional to the exchange interaction in a spin
pair. It is obvious that light polarized perpendicularly to the direction of
the chain should be absorbed. The extent of absorption continuum will be
determined by the extent of the magnetic-excitation spectrum and the
intensity will be determined by the density of states.

For a uniform Heisenberg chain of $S=$1/2 spins, the magnetic-excitation
spectrum extends from 0 to $\nu_m=\pi J$ (if only nearest-neighbor
interaction is taken into account)~\cite{c10}. Taking $J=560$\,K for
NaV$_2$O$_5$~\cite{c4}, we find that at $T>T_{\rm sp}$ the absorption
continuum extends over approximately 2400\,cm$^{-1}$. At $'<T_{\rm sp}$ the
magnetic atoms in the chain dimerize, the chain becomes alternating, and a
gap appears in its magnetic-excitation spectrum~\cite{c11}. According to the
inelastic neutron scattering data for NaV$_2$O$_5$, the smallest gap opens at
the point {\bf q}=(1,1/2,0) (the fact that the gap is not located at the
center of the BZ is a consequence of the interchain interaction) and its
width is $\Delta$=9.8\,meV=79\,cm$^{-1}$~\cite{c5}. In this case the
continuum of two-magnon excitations at $T<T_{\rm sp}$ starts above
$2\Delta=158$\,cm$^{-1}$. A feature due to the density of states near the gap
can be expected to appear at the frequency $\nu=2\Delta$.

The behavior of the absorption continuum in the IR spectra agrees
qualitatively with the picture described above: The continuum is observed
only in {\bf E}$||${\bf a} polarization, and below the SP transition
temperature the intensity of its long-wavelength part decreases.  At
$T<T_{\rm sp}$, a weak line is present at the frequency
$\nu$=157.2\,cm$^{-1}\simeq 2\Delta$. However, it does not shift as
$T\to T_{\rm sp}$, so that it cannot be attributed to a gap-associated
feature.

The interaction of the lattice vibrations with the magnetic-excitation
continuum is manifested in the asymmetric line shape that is characteristic
for a Fano resonance~\cite{c6}. Such a line against the continuum background
is shown in Fig.\,2 for temperatures 101\,K, 37\,K$>T_{\rm sp}$. As a result
of the change in the magnetic excitation spectrum at the SP transition, the
spin-phonon interaction and hence the line shapes change. The most striking
manifestations of this effect should be expected at low frequencies, where
below $T_{\rm sp}$ the continuum vanishes on account of the opening of the
gap in the magnetic-excitation spectrum. This vanishing of the continuum at
$T<T_{\rm sp}$ and the associated transformation of the line shape from a
Fano dispersion contour to an ordinary symmetric absorption contour are
clearly seen in Fig.\,2.

A more detailed investigation of the effects observed requires data on the
dispersional dependencies of the phonon and magnetic-excitation energies in
NaV$_2$O$_5$. Such data do not exist at present.

We thank E.A.Vinogradov, O.N.Kompanets, and G.N.Zhizhin for supporting this
work. This work was made possible in part by
Grants Nos. 95-02-03796-a and 96-02-19474 from
the Russian Fund for Fundamental Research

\newpage

Fig.\,1. Transmission spectrum of a  NaV$_2$O$_5$ single crystal
for two polarizations of the light,
${\bf E}||{\bf a}$ (top) T=15\,K$<T_{\rm sp}$ --- solid line,
T=38\,K$<T_{\rm sp}$ --- dotted line;
${\bf E}||{\bf b}$ (bottom) T=15\,K$<T_{\rm sp}$ --- solid line,
T=42\,K$<T_{\rm sp}$ --- dotted line. Spectral resolution: 1\,cm$^{-1}$.

Fig.\,2. Line near 90\,cm$^{-1}$ in ${\bf E}||{\bf a}$ polarization
at temperatures 101 and 37\,K$>T_{\rm sp}$ and 6\,K$<T_{\rm sp}$.
The spectra were recorded with resolutions (top to bottom) of
0.05, 0.2, and 1\,cm$^{-1}$.

Fig.\,3. Temperature dependence of the integrated intensity of a line near
101\,cm$^{-1}$ in ${\bf E}||{\bf b}$ polarization. Solid line is a guide for
the eye.
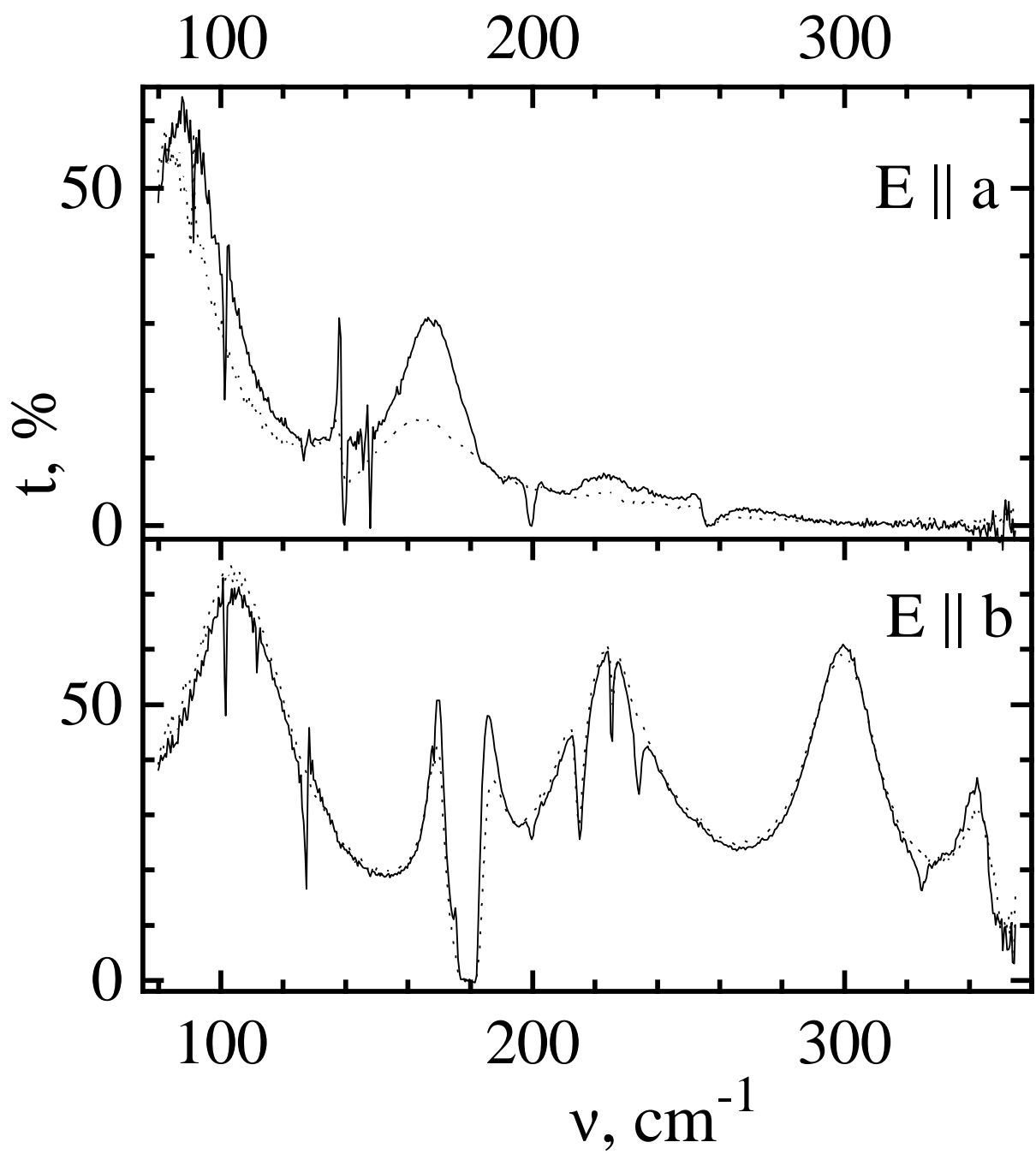
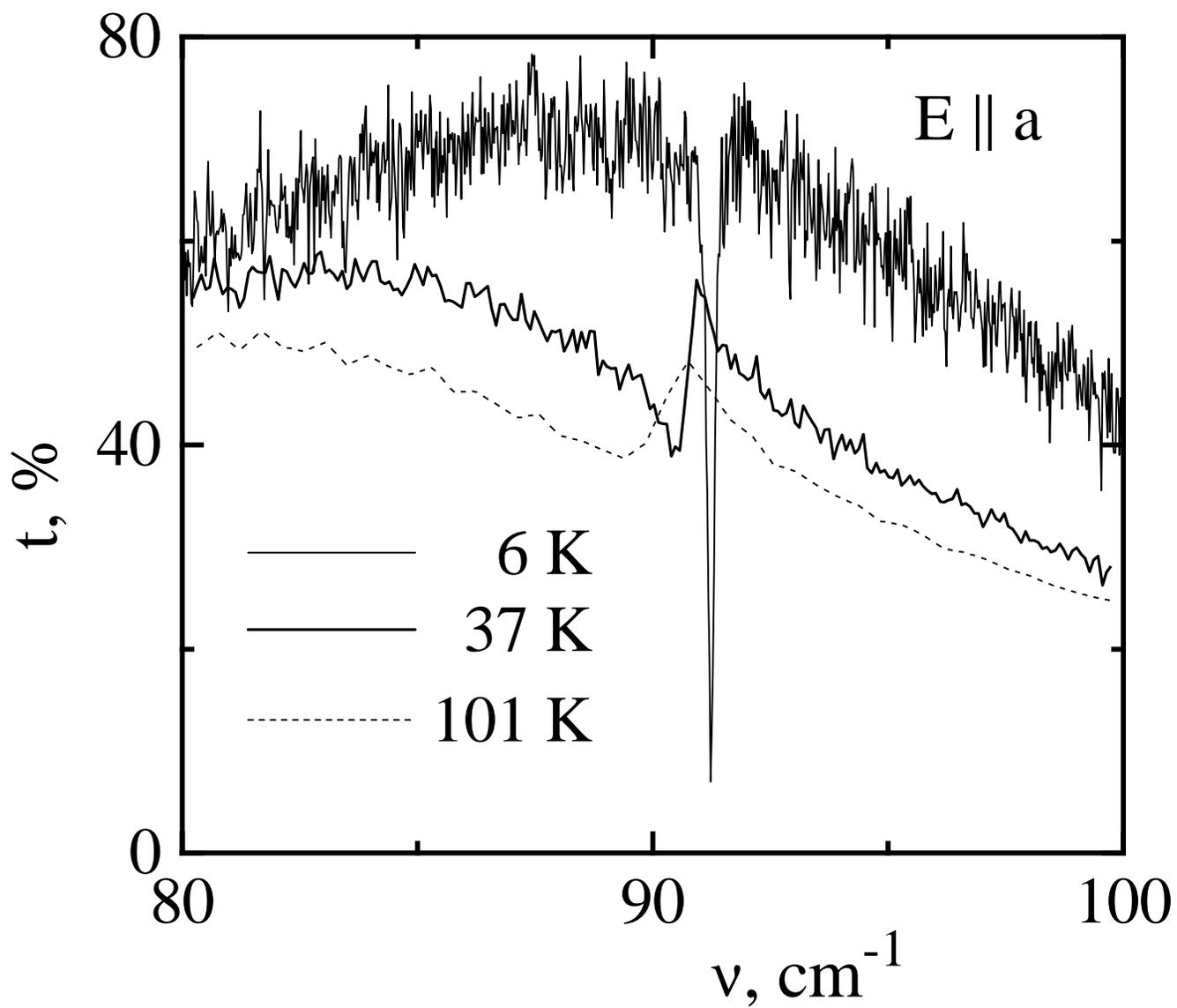
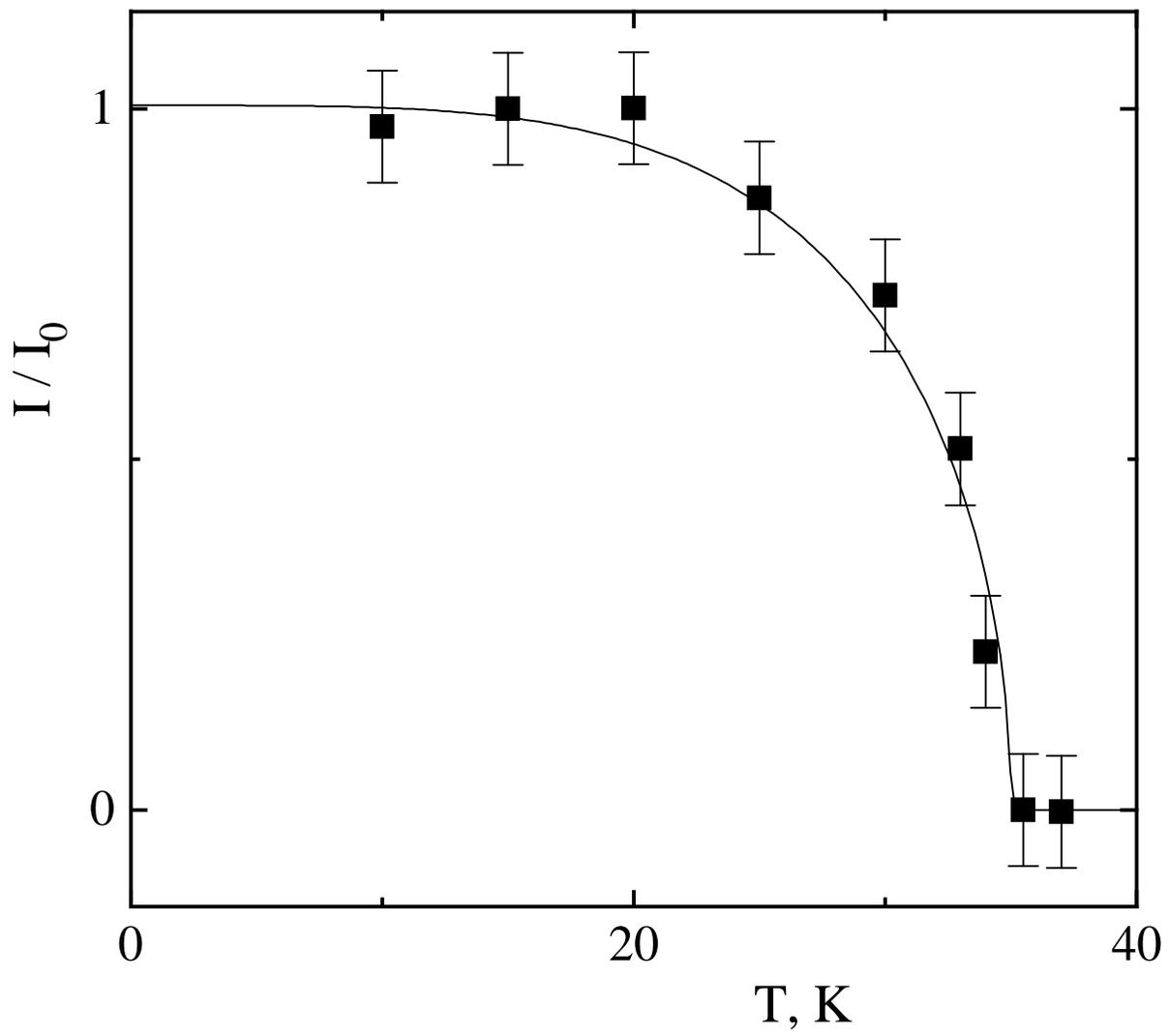
\end{document}